\documentclass[%
 reprint,
superscriptaddress,
 amsmath,amssymb,
 aps,
]{revtex4-2}
\usepackage{siunitx}
\usepackage{graphicx}  % Include figure files
\usepackage{subfigure}
\usepackage{multirow}
\usepackage{tabularx}
\usepackage{fancyhdr}
\usepackage{longtable}
\usepackage{parskip}
\usepackage[T1]{fontenc}
\usepackage{dcolumn}   % Align table columns on decimal point
\usepackage{gensymb}
\usepackage{bm}        % bold math
\usepackage{amsfonts}  % Common math fonts
\usepackage{amsmath}   % Common math functions
\usepackage{amssymb}   % Common math symbols
\usepackage{hhline}
\usepackage{xcolor}
\usepackage{ulem}
\usepackage{gensymb}
\usepackage{soul}
\begin{document}

\preprint{APS/123-QED}

\title{Evidence for fully-gapped superconductivity in BCS superconductor NiBi$_{3}$\\}% Force line breaks with \\
%\thanks{A footnote to the article title}%

\author{Atul Gangwar}
\affiliation {\it Department of Physics, Indian Institute of Technology (BHU), Varanasi 221005, India}

\author{Kunal Yadav}
\author{S.Patnaik}
\affiliation {\it School of Physical Sciences, Jawaharlal Nehru University, New Delhi 110067, India
}

%\author{S.Patnaik}
%\affiliation {\it School of Physical Sciences, Jawaharlal Nehru University, New Delhi 110067, India}

\author{Sandip Chatterjee}
\email{schatterji.app@itbhu.ac.in}
\affiliation {\it Department of Physics, Indian Institute of Technology (BHU), Varanasi 221005, India}

%\date{\today}% It is always \today, today,
             %  but any date may be explicitly specified

\begin{abstract}
We investigate the physical characteristics, normal state and superconducting properties of NiBi$_3$ single crystals. Measurements of electrical resistivity, magnetization, and London penetration depth demonstrate a superconducting transition temperature, $T_c \sim 4.0$ K, with a sharp transition in resistivity ($\Delta T_c = 0.20$ K) and RRR = 18, reflecting the high quality of NiBi$_3$ single crystals.
With both orientations H$\perp$b and H$\parallel$b of NiBi$_3$ single crystals, we estimated the upper critical field, $H_{c2}$, from the magnetization data.  In both orientations, $H_{c2}$ is significantly smaller than the Pauli limit, suggesting the orbital pair breaking in superconducting state. The coherence length, $\xi_{GL}(0) = 26.78$ nm, electron-phonon coupling constant, $\lambda_{\text{e-ph}} = 0.81$, and penetration depth, $\lambda_{GL}(0) = 181.8$ nm, suggest type-II superconductivity in NiBi$_3$. In the superconducting state, $\Delta \lambda$(T) is best described by an $s$-wave BCS model and does not have linear or quadratic dependency with T, which is in line with the expectation for a nodeless superconducting order parameter. The temperature evolution of the superfluid density, obtained from $\lambda (T)$, reveals a fully gapped superconductivity in NiBi$_3$, with a superconducting gap $\frac{2\Delta_0}{k_BT_c}=4.07$. All the results from the present study indicate that NiBi$_3$ is a typical type-II, BCS-like, moderately strong coupled, and fully gapped superconductor in the dirty limit.       
\end{abstract}

%\keywords{Suggested keywords}%Use showkeys class option if keyword
                              %display desired
\maketitle

%\tableofcontents

\section{\label{sec:level1}	INTRODUCTION }

Topological superconductors (TSCs) have drawn growing interest in the topological family because of their unique property to host Majorana fermions and fault tolerant quantum computing, and hence become the most promising materials in scientific community to search for the topological superconductivity \cite{TSC, TSC1}. Moreover, superconductors exhibiting unconventional pairing symmetries have gained a center of attention, as they give rise to novel quasiparticle excitations and potential discovery of new quantum phases of matter. It is well known that spin orbit coupling (SOC), often arising from the presence of heavy atoms and can give rise to various intriguing phenomena like nontrivial electronic topology and spintronic effects. Among all the heavy elements that exhibit SOC, bismuth (Bi) has emerged as one of the most widely studied and promising candidates from p-block \cite{Ti/TSC}. Bi is a strong topological insulator. Therefore, it is expected that studies of compounds containing bismuth will keep attracting attention of researchers focused on discovery of new topological phases \cite{TIIS, NPJNMRBi2Se3Te3}.
The discovery of first experimental TSC, Cu$_x$Bi$_2$Se$_3$ \cite{CuBi2Se3} has generated considerable attention, as this system has been widely proposed as a promising platform for realizing topological superconductivity.    

Intermetallic compounds formed from two or more metallic elements of d and p-block with well-defined stoichiometric ratios and distinct crystal structures \cite{Prbnibi,Prmnibi}, have been extensively studied in condensed matter physics because of their abundance and unique chemical and physical properties. Among the elements commonly found in such compounds, Bi plays a particularly interesting role due to its unique electronic structure and bonding behavior \cite{Nibilayersciadv,APLnibilay,PhysRevResearchnibilayer}, since Bi is a strong topological insulator with non-vanishing weak indices and also it introduces some covalent character to the bonding. NiBi$_3$ primarily exhibits metallic bonding due to the interaction between nickel (Ni) and bismuth (Bi) atoms. However, the presence of bismuth, which possesses strong spin-orbit coupling (SOC), can be considered one of the key factors contributing to the emergence of exotic topological phases in a superconductor NiBi$_3$ \cite{NiBiARPES,NiBitheory} because in these types of alloys, Ni generally tends to lose its magnetic moment \cite{prmnibilayer,BHATnibifilm,scnibi3jpsjapan}.

NiBi$_3$ exhibits superconducting transition temperature $T_{c} \sim 4.10$~K and crystallizes in centrosymmetric orthorhombic structure, P\texttt{nma} space group. It is considered a type-II conventional superconductor within the framework of BCS (Bardeen–Cooper–Schrieffer) theory.
In previous works, some researchers \cite{NiBiARPES,NiBifullygap,scnibi3jpsjapan,Nibi3VPS} demonstrated that ferromagnetism is absent in bulk single crystals, while others \cite{2mag2012,3mag2013,4magESR,NiBi2026} reported the coexistence of superconductivity and ferromagnetism, evidenced by  isothermal hysteresis in magnetization measurements\cite{3mag2013}. The Ni-Bi phase diagram clearly indicates that, in the $1:3$ stoichiometry, avoiding magnetic impurities during the synthesis process is highly challenging. In previous studies, the primary explanation for the observed ferromagnetism was the presence of residual Ni inclusions within the crystals, which could contribute to the measured magnetic moments\cite{4magESR}. Since Bi does not exhibit superconductivity within the investigated temperature range, any remanent Bi flux on the surface of the single crystals is expected to have a negligible impact on their superconducting properties\cite{nibi3scrystSCTech2018}. In our work, we used flux method to synthesize high-quality single crystals of NiBi$_3$. These crystals exhibit a very sharp superconducting transition, indicating morphological homogeneity. Additionally, the intrinsic strong electron scattering in this material results in extremely type-II superconductivity. Low-temperature electrical resistivity and specific heat measurements suggest that a fully-gapped BCS model best describes the superconducting order parameter of NiBi$_3$.

\section{\label{sec:level2}	    EXPERIMENTAL PROCEDURE}

Single crystals of the intermetallic compound NiBi$_3$ were grown by self-flux method in excess of the Bi. High-purity Bi pieces (Alfa Aesar,99.999\%) and Ni powders (Alfa Aesar, 99.999\%) were loaded into the alumina crucible in the ratio of Ni:Bi = 1:15. The crucible was sealed in quartz ampoule under high vacuum up to $10^{-6}$ mbar. The evacuated quartz ampoule was heated in Muffle furnace with a rate of 100\textdegree C/h, up to 1050\textdegree C, and then maintained at this temperature for 24 hours to get the better homogeneity. Afterward, it was slowly cooled down to 380\textdegree C at a rate of 2\textdegree C/h. The ampoule was then taken out of the furnace and rapidly centrifuged to remove the Bi flux. We obtained high quality, needle-like single crystals, with the largest dimensions: $\sim 10mm \times 1mm \times 1mm$, as shown in the inset of figure $1$, a photograph of single crystals of NiBi$_3$ placed on a millimeter paper. 

The phase purity and crystal structure of the sample were examined using X-ray diffraction (XRD). Room temperature XRD measurements were performed on the powdered sample using Cu-K$_\alpha$ radiation on a Rigaku HR-XRD diffractometer.
Magnetic susceptibility measurements were carried out in a quantum design magnetic property measurement system (QD MPMS 3).
Resistivity was measured using a quantum design physical properties measurement system ( QD Dynacool PPMS), using a standard four-probe method to make the contacts. Penetration depth measurements were carried out in a cryogen free magnetic system (CFMS) using a tunnel diode oscillator (TDO).

\begin{figure}[h!]
\centering
\includegraphics[width=\columnwidth]{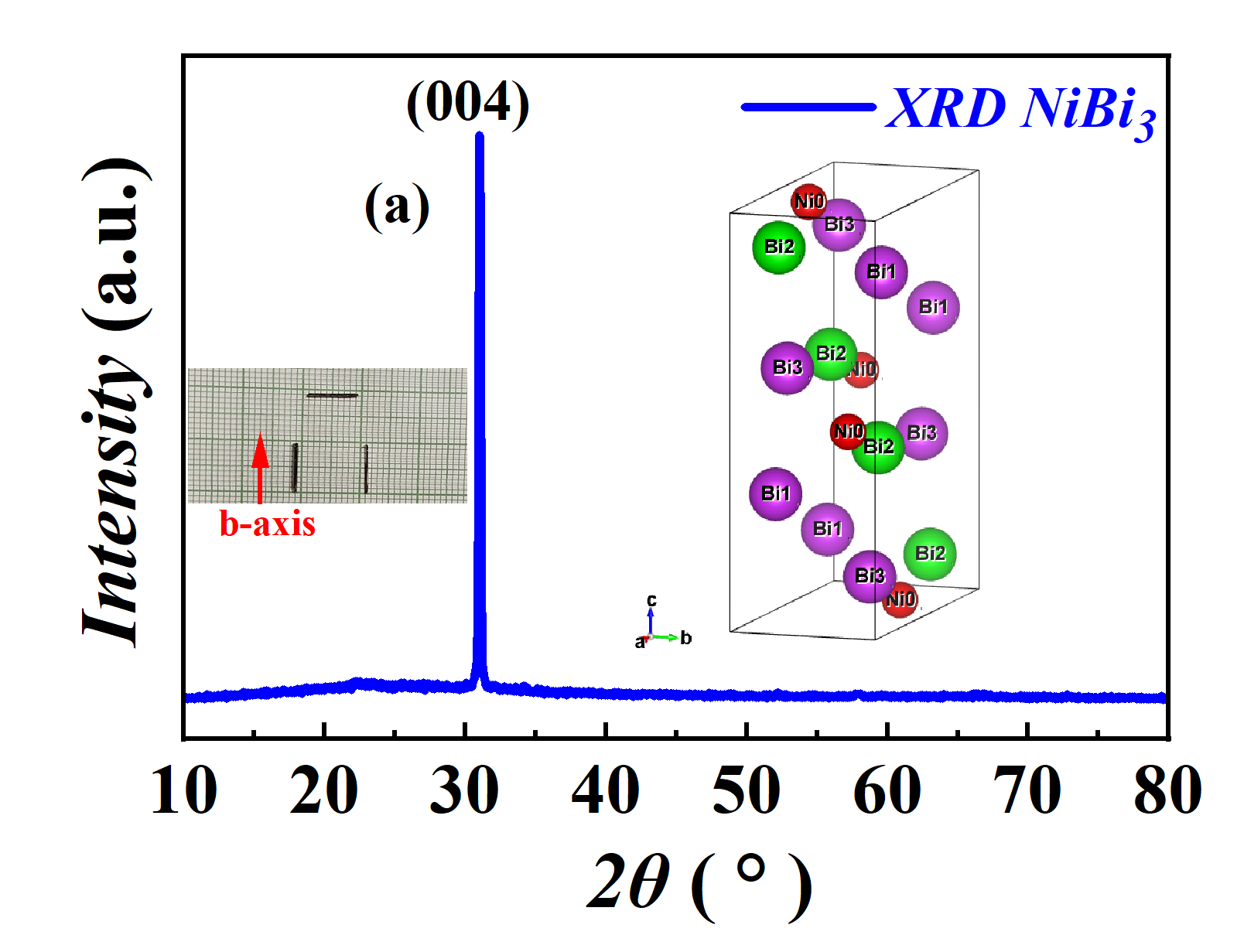}
\caption{shows the single crystal reflection of NiBi$_3$. The left inset shows a photograph of single crystals NiBi$_3$ on a millimeter paper and the right inset shows the unit cell corresponds to Ni and Bi atoms.}
\label{fig:XRD nibi}
\end{figure}

\section{\label{sec:level3}	 RESULTS AND DISCUSSIONS}

\subsection{\label{sec:level4} X-Ray Diffraction}
Figure (1) shows the XRD pattern of needle like single crystal of NiBi$_3$, displaying only (00L) reflections which indicates that crystallographic c-axis is oriented perpendicular to the sample surface\cite{NiBifullygap}. There is no impurity peaks present in spectra, indicates the high purity of the crystal.

\begin{figure*}[t]
    \centering
    \includegraphics[width=1\textwidth]{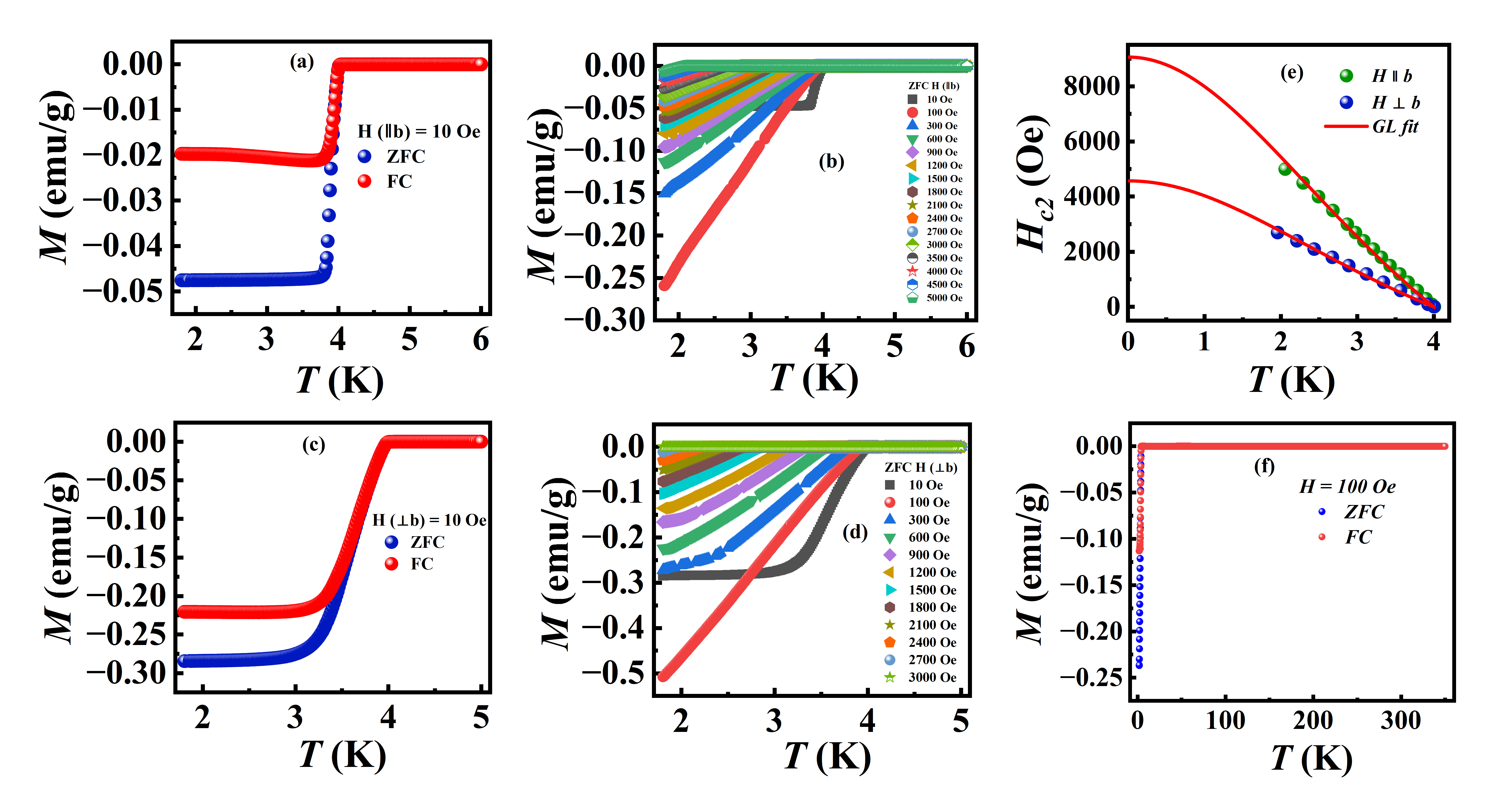}
\caption{ (a)-(d) temperature dependent DC magnetization with field applied perpendicular and parallel to the $b$ direction. (e) represents the $H_{c2}$ phase diagrams for both directions ($H \parallel b$) and ($H \perp b$), solid red lines are the generalized GL fit. (f) shows magnetization in temperature range 1.8-350 K.}
\label{fig:MTallnibi}
\end{figure*} 

\subsection{\label{sec:level5} Magnetization}
The superconductivity of NiBi$_3$ was first confirmed by study of DC magnetization in zero field cooled (ZFC) and field cooled (FC) modes under an applied field of $10$ Oe in both the directions H $\perp$ b and H $\parallel$ b of the crystal as shown in figures 1 (a) and (c), a clear superconducting transition is found at 4 K and bifurcation in both the curves suggesting type-II superconductivity. To check the magnetic impurity due to Ni inclusions as reported by some previous groups \cite{2mag2012,3mag2013,4magESR,NiBi2026} we also performed magnetization in temperature range 1.8 - 350 K at 100 Oe magnetic field in both ZFC and FC modes, figure 1(f) which indicates that sample is completely diamagnetic above the superconducting transition through out the temperature range. To calculate the second critical field (H$_{c2}$) and for greater consistency in determining the H$_{c2}$, we performed magnetization measurements in both the directions H $\parallel$ b and H $\perp$ b of crystal as shown in figure 1 (b) and (d) respectively.  We noted down the points $T^{\text{onset}}_c$ corresponds to the temperature above which system enters in the diamagnetic regime.
 
\begin{equation}
H_{c2}(0) = -0.693 \, T_c \left[ \frac{dH_{c2}}{dT} \right]_{T=T_c}  
\label{eq:Hc2}
\end{equation}
We fitted our data using the generalized Ginzburg-Landau model,
\begin{equation} 
H_{c2}(T) = H_{c2}(0) \frac{1 - \left(\frac{T}{T_c}\right)^2}{1 + \left(\frac{T}{T_c}\right)^2}
\label{eq:Hc2}
\end{equation}

In the previous reports \cite{2mag2012,4magESR}, the Werthamer,Helfand and Hohenberg (WHH) model fitting was performed at midpoint temperature  ($T^{\text{mid}}_c$) using equation (1). This model describes orbital limited $H_{c2}$ of type-II superconductors in the dirty-limit. This fit obviously underestimates the values of $H_{c2}$, and it does not reflect the positive upturn that becomes more noticeable at lower temperatures.  
The figures $2$(a) shows the phase diagrams of $H_{c2}(T)$. In the main panel of figure $2$(e), the solid red line represents the extrapolated Ginzburg–Landau (GL) fitting for $T^{\text{onset}}_c$ obtained from M(T) with fields in H $\parallel$ b and H $\perp$ b directions respectively. As shown figure 2(e), the solid red line fits well with the GL-equation. The estimated values of  $H_{c2}(0)$ is $9058.23\pm 100.85$ Oe for H $\parallel$ b and $4590.33\pm 57.29$ Oe for H $\perp$ b. 
 These values are comparable to previous results \cite{2mag2012,3mag2013,4magESR,NiBifullygap}, while remaining significantly smaller than the Pauli limit, i.e., $1.86T_C \approx 7.5 T$. The excellent agreement of the data with the Ginzburg–Landau model suggests the conventional superconductivity, indicating that orbital pairing mechanism is dominant in NiBi$_3$.
 The observed anisotropy in $H_{c2}$ is highly characteristic of an anisotropic GL superconductor. In our needle-like single crystals, electronic transport and metallic bonding are expected to be stronger along the longitudinal growth axis, resulting in an anisotropic coherence length where the longitudinal component exceeds the transverse component ($\xi_\parallel > \xi_\perp$). Because the upper critical field is inversely proportional to the area enclosing the superconducting vortex core, the parallel configuration is restricted by the smaller transverse coherence length.
 In addition, we used upper critical field to calculate the Ginzburg-Landau coherence length, $\xi$(0), for both orientation using formula given below \cite{BCStheory}, 
\begin{equation}
H_{c2}(0) = \frac{\Phi_0}{2\pi \xi(0)^2}
\end{equation}
Here $\Phi_0 = 2.067 \times 10^{-15} \, \text{Wb}$ is the flux quantum. The estimated coherence lengths are $\xi_{\parallel b}(0) = 19.08$ nm and $\xi_{\perp b}(0) = 26.78$ nm, yielding an isotropy ratio $\gamma = \frac{\xi_{\parallel b}}{\xi_{\perp b}} \approx 1.97$. The resulting anisotropy ratio of $\gamma$ reflects the intrinsic electronic and structural anisotropy. All these values strongly indicate type-II moderately coupled superconductivity in NiBi$_3$. 

\begin{figure}[h!]
\centering
\includegraphics[width=\columnwidth]{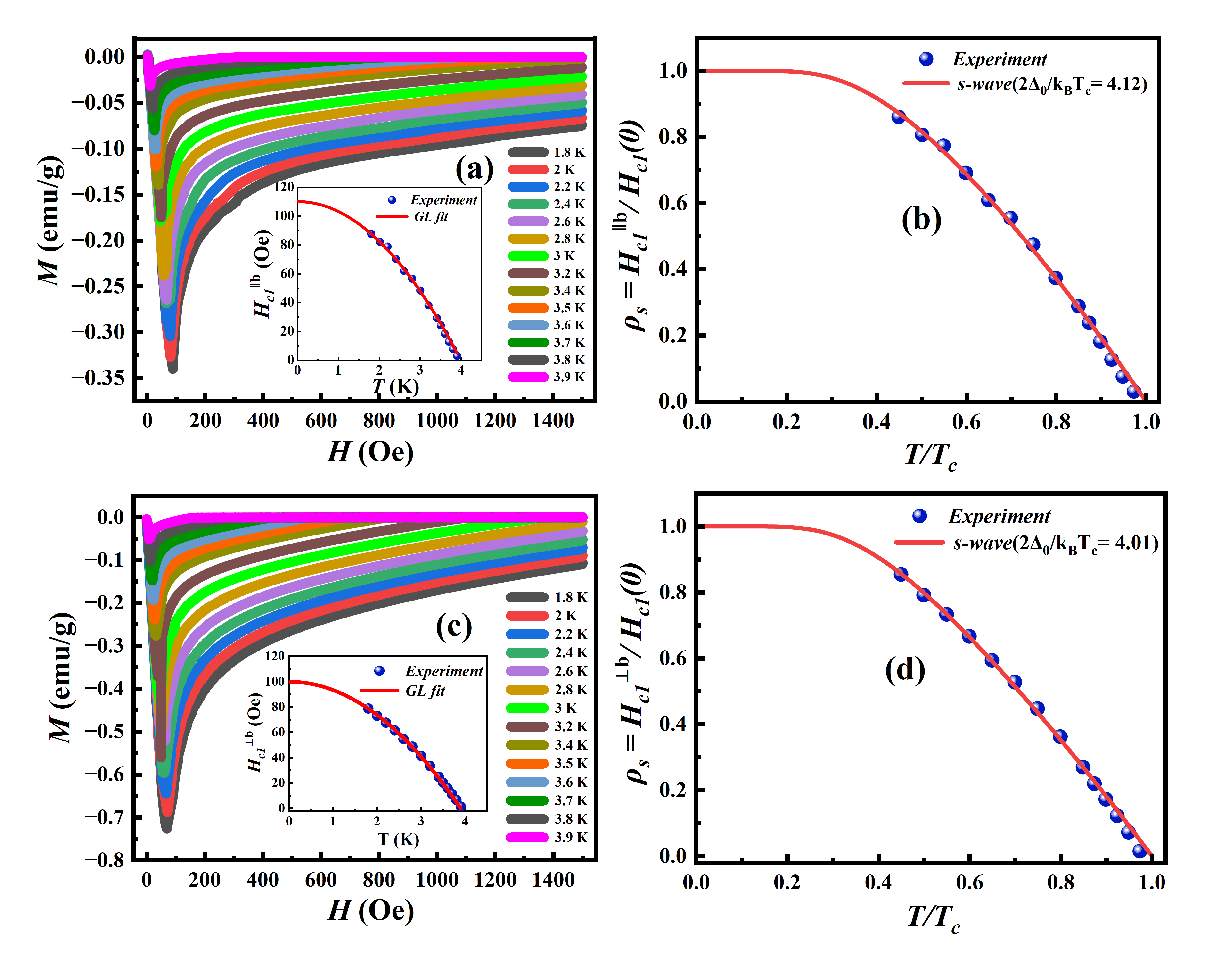}
\caption{ (a) and (c) show M(H) isotherms at various temperatures for both parallel and perpendicular directions respectively. (b) and (d) shows the calculated superfluid density ($\rho_s$) from $H_{c1}$ and solid red lines are BCS s-wave fittings.}
\label{fig:Hc1NiBi3}
\end{figure}

Figure (3) shows field dependent magnetization curves M(H) at different temperatures. The lower critical field ($H_{c1}$) was determined from the initial isothermal M(H) curves at low fields where M(H) deviates from the initial Meissner state. In the perfect Meissner state, the sample behaves as a perfect diamagnet, yielding a linear relation between the applied field ($H_{app}$) and the magnetization. 
The lower critical field $H_{c1}$ was estimated using the Ginzburg Landau model \cite{BCStheory},
\begin{equation}
   H_{c1}(T) = H_{c1}(0)\left[1 - \left(\frac{T}{T_c}\right)^2\right]
\end{equation}
as shown in the inset of figure 3 (a) and (c) for H $\parallel$ b and H $\perp$ b directions respectively. The obtained value of $H_{c1}$ are 110.08 Oe for H $\parallel$ b and 99.82 Oe for H $\perp$ b. The penetration depth $\lambda$ can be determined using the relation $\lambda = \kappa .\xi$, where $\kappa = \sqrt{\frac{H_{c2}(0)}{H_{c1}(0)}}$. For H $\parallel$ b, $\kappa = 9.07$ and $\lambda (0) = 172.8$ nm while for H $\perp$ b, $\kappa = 6.79$ and $\lambda (0) = 181.8$ nm. The values of $\kappa$ is greater than $1/\sqrt{2}$, indicating the type-II superconductivity in NiBi$_3$.

The temperature dependence of $H_{c1}$ provides a powerful macroscopic window into the microscopic physics of the superconducting state. Because $H_{c1}$(T) scales inversely with the square of London penetration depth ($H_{c1} \propto \lambda^{-2}$) and serves as a direct probe of stiffness of superconducting condensate.
To extract the normalized superfluid density ($\rho_s (T)$), we utilized the London relation $\rho_s (T) = H_{c1}(T)/H_{c1}(0)$, where $H_{c1}(0)$ is the lower critical field at T = 0 K.

For an isotropic, single-gap $s$-wave superconductor, the temperature dependence of the normalized superfluid density $\rho_s(T)$ is described by the standard weak-coupling BCS theory as \cite{BCStheory,M.Tinkham}, 

\begin{equation}
\rho_s(T) = 1 - 2 \int_{\Delta(T)}^{\infty} \left( -\frac{\partial f}{\partial E} \right) \frac{E}{\sqrt{E^2 - \Delta(T)^2}} \, dE
\end{equation}

Where:
    $f(E) = \frac{1}{e^{E/k_B T} + 1}$ is the Fermi-Dirac distribution function. $E$ is the quasiparticle excitation energy.
$\Delta(T)$ is the temperature-dependent superconducting energy gap given as,
\begin{equation}
\Delta(T) = \Delta(0) \tanh \left[ 1.82 \left( 1.018 \left( \frac{1}{t} - 1 \right) \right)^{0.51} \right]
\end{equation},

with the reduced temperature $t = T/T_c$. For weak-coupling $s$-wave BCS superconductors, the zero-temperature gap scales universally with the critical transition temperature $T_c$, as $\Delta(0) = 1.764 \, k_B T_c$ or $2\Delta(0)/k_B T_c = 3.532$.
Figure 3(b) and (d) represents the normalized $\rho_s(T)$ for H $\parallel$ b and H $\perp$ b directions respectively. We fitted the data with different theoretical models but it nicely fitted the s-wave, suggesting a fully gapped superconductivity. The solid red lines are the single gap s-wave fit. The fitting yields $2\Delta(0)/k_B T_c = 4.12$ for  H $\parallel$ b and  $2\Delta(0)/k_B T_c = 4.01$ for H $\perp$ b, which lies in the dirty limit of BCS prediction.

\subsection{\label{sec:level7} Resistivity}
\begin{figure}
    \centering
    \includegraphics[width=0.9\linewidth]{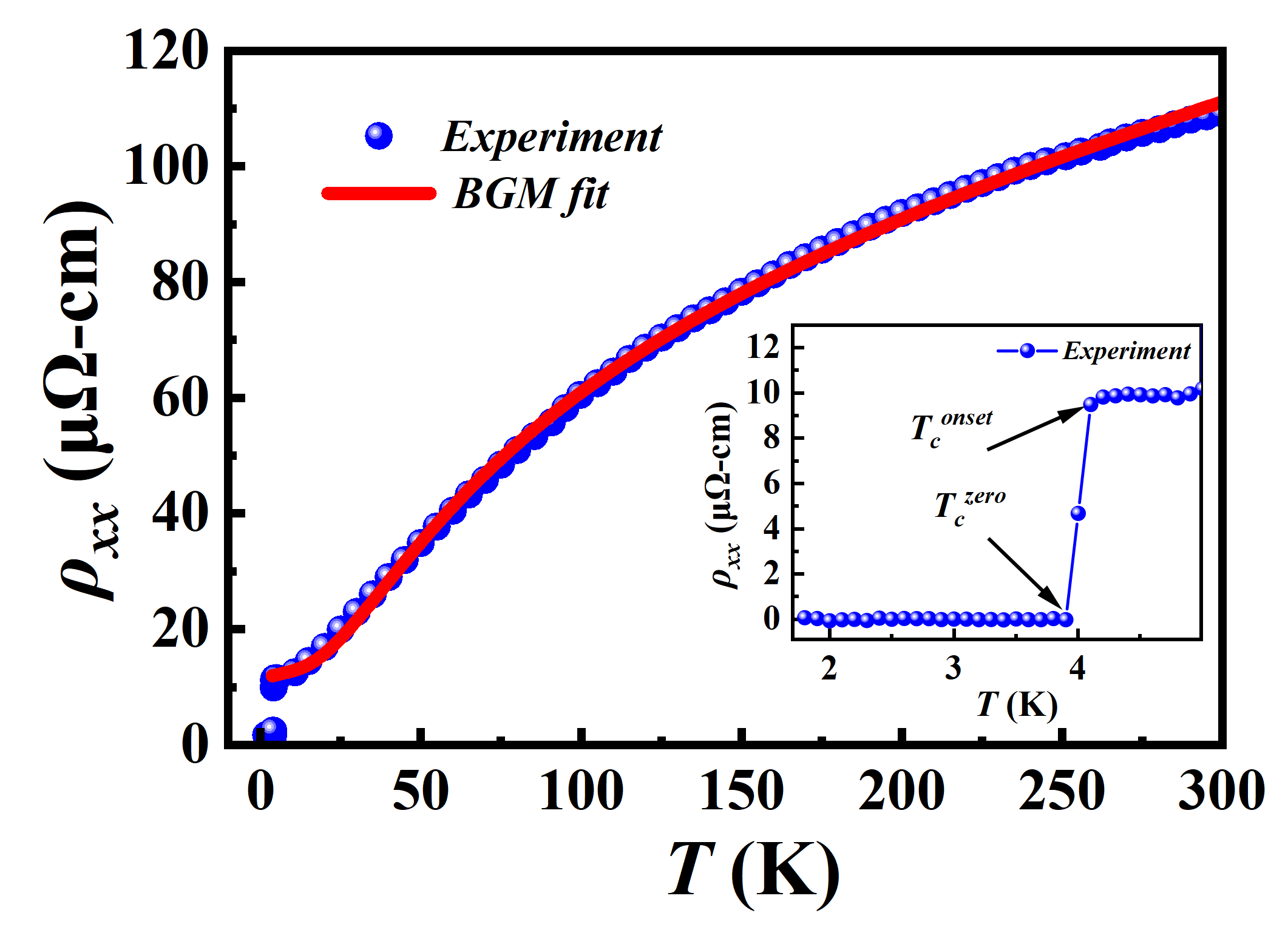}
    \caption{shows the temperature-dependent electrical resistivity $\rho_{xx}$ in the temperature range  between $2$ and $300$ K in zero applied magnetic field and solid red line is the Bloch Gruneisen-Mott (BGM) fitting of equation (7) as discussed in text}.
    \label{fig:RTNiBi3}
\end{figure}

The temperature evolution of the longitudinal resistivity ($\rho_{xx}$) of NiBi$_3$ in the temperature range from $2$ - $300$ K is shown in figure 4. The $\rho_{xx}$ reveals a typical metallic behavior with no sign of anomaly in the normal state. The low temperature regime, as shown in the inset of figure 4, which clearly displays the superconducting transition, with $T^{\text{onset}}_c = 4.1$ K and $T^{\text{zero}}_c = 3.9$ K. The former is the temperature below which $\rho_{xx}$ starts deviating from the normal-state value upon cooling. The latter is selected at the temperature where $\rho_{xx}$ drops to zero. The fact that the residual resistivity ratio RRR, defined as $ \frac{\rho(300 \, \text{K})}{\rho_{0}}$, amounts to 18, and the sharp superconducting transition with the width $\Delta T_c \left( T^{\text{onset}}_c - T^{\text{zero}}_c \right) \sim 0.20$~K, suggests the high quality of the single crystals.
We fit the $\rho_{xx}(T)$ using a Bloch-Gruneisen (BG) formula with an additional Mott term as \cite{C.Kittel}, 
$\rho_{xx}(T) = \rho_{0} + \rho_{BG}(T) + \rho_{Mott}(T)$. where
 \begin{equation}
\rho_{BG}(T) = 4R \left( \frac{T}{\Theta_{\rm D}} \right)^5 \int_0^{\Theta_{\rm D}/T} \frac{t^5}{(e^t - 1)(1 - e^{-t})} \, dt
\end{equation} 
and $\rho_{Mott}(T)=-\alpha T^3$,
where $\rho_{0}$ is the residual resistivity. 
The second term describes the electron-phonon scattering, where $\theta{_D}$ is the Debye-temperature and $R$ is coupling constant. The third term describes the $s$-$d$ interband scattering where $\alpha$ is the Mott coefficient\cite{strangeresitNibi3}. The fitting yields $\rho_{0}$ = $11.46\pm 1.02\, \mu\Omega\text{cm}$, $R$ = $78.02\pm 1.39\, \mu\Omega\text{cm}$, $\theta_{D}$ = $105.20\pm 2.82$~K which is consistent with the value obtained in previous study and $\alpha$ = $3.24\pm0.08 \times 10^{-6}\,
  \mu\Omega\text{cm}\,\text{K}^{-3}$.
These values are consistent with the ones reported in Ref.~\cite{NiBifullygap,strangeresitNibi3,Nibi3VPS}.

\subsection{\label{sec:level6} Penetration Depth and Superfluid Density}

\begin{figure*}[t]
    \centering
    \includegraphics[width=1\textwidth]{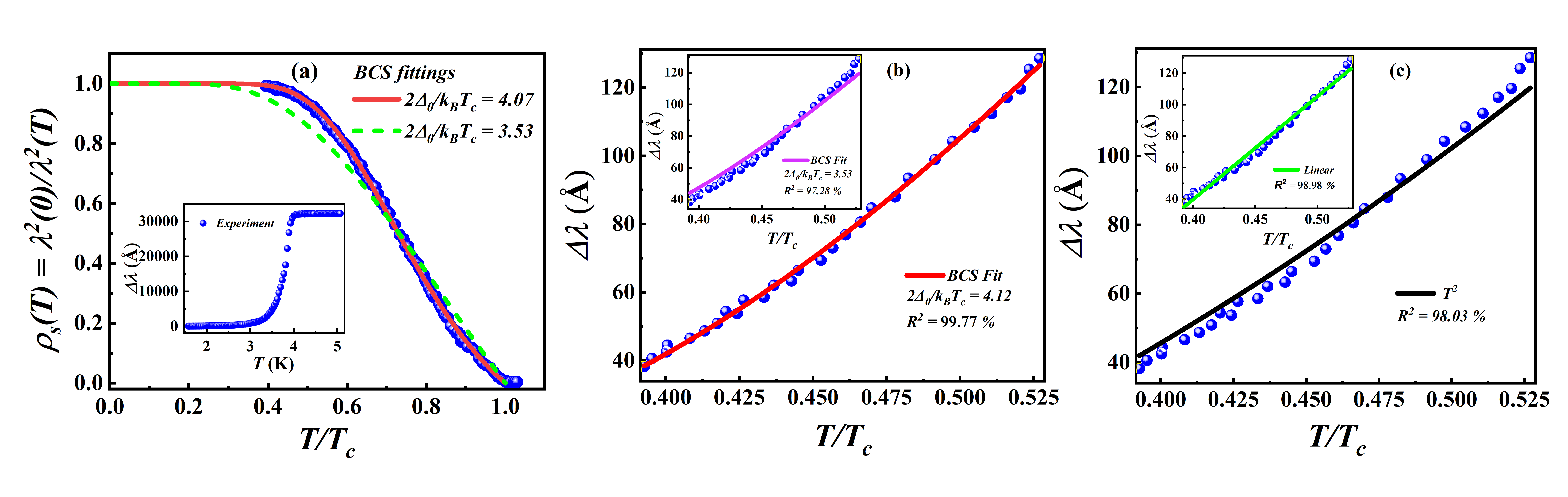}
\caption{(a) shows the superfluid density calculated from penetration depth $\lambda (T)$ and solid red line is the BCS fitting with $\frac{2\Delta_0}{k_B T_c}$ = 4.07 while green dotted line is the standard BCS fitting. Inset shows the variation of $\Delta \lambda$ with T. (b) shows the low temperature penetration depth $\Delta  \lambda (T)$ as a function of reduced temperature (T/$T_c$), fitted with exponential model (solid red line with $\frac{2\Delta_0}{k_B T_c}$ = 4.12) and the upper inset shows the BCS model fitting with standard $\frac{2\Delta_0}{k_B T_c}$ = 3.53. (c) shows $T^2$ fitting with $R^2 = 98.03 \%$ and the upper inset shows the linear fitting with $R^2 = 98.98 \%$.}
\label{fig:penetraNiBi3}
\end{figure*} 

To understand the superconducting pairing symmetry in NiBi$_3$, the temperature dependence of the magnetic penetration depth, $\Delta\lambda(T)$, was measured using the tunnel diode oscillator (TDO) technique. This technique is highly sensitive to small changes in the resonant frequency and is widely used to study the superconducting gap structure. The measured frequency shift, $\Delta F(T)$, was converted into the change in penetration depth using the calibration factor $G=2.27~\mathrm{\AA/Hz}$ for our experimental setup\cite{v5ks-l9qs}. The obtained penetration depth data are shown in inset of figure 5(a). A sharp increase in $\Delta\lambda(T)$ is observed near the superconducting transition temperature, $T_c\approx4.0$ K, indicating a high-quality sample with a well-defined superconducting transition. At low temperatures, thermal excitations are very small, making this region suitable for investigating the superconducting gap symmetry.

The low-temperature data were first fitted using the conventional isotropic BCS model, which predicts an exponential temperature dependence of the penetration depth \cite{BCStheory,M.Tinkham},

\begin{equation}
\Delta\lambda(T)=\lambda(0)\sqrt{\frac{\pi\Delta_0}{2k_BT}}
\exp\left(-\frac{\Delta_0}{k_BT}\right),
\end{equation}

where $\Delta_0$ is the superconducting energy gap at zero temperature and $\lambda(0)$ is the zero-temperature penetration depth obtained from the upper critical field analysis. As shown in figure 5 (b), the BCS model fits the experimental data very well, giving a coefficient of determination of $R^2=99.77\%$. The fit gives a superconducting gap ratio of $\frac{2\Delta_0}{k_BT_c}=4.12$, which is higher than the weak-coupling BCS value of 3.53 (inset of fig.5(b)). This result suggests that NiBi$_3$ is a moderately strongly coupled superconductor with a fully gapped $s$-wave superconducting gap.

To check whether the superconducting gap contains line nodes, the same data were also fitted using a linear temperature dependence,

\begin{equation}
\Delta\lambda(T)\propto T,
\end{equation}

which is expected for clean $d$-wave superconductors. The result is shown in the inset Fig.5(c). The linear fit gives $R^2=98.98\%$, which is lower than the BCS fit. In addition, clear deviations from the experimental data are observed. Therefore, the penetration depth does not follow the linear temperature dependence expected for a line-nodal superconducting gap.

The data were further analyzed using a quadratic power-law dependence,

\begin{equation}
\Delta\lambda(T)\propto T^2,
\end{equation}

which is often associated with point nodes or impurity-scattered nodal superconductors. As shown in Fig.5(c), the fit gives $R^2=98.03\%$, which is also lower than the BCS fit. Thus, both the linear and quadratic power-law models fail to describe the low-temperature penetration depth as accurately as the exponential BCS model. These results provide strong evidence that NiBi$_3$ is a fully gapped superconductor with conventional $s$-wave pairing symmetry. The larger value of $2\Delta_0/k_BT_c$ further indicates that superconductivity in NiBi$_3$ lies in the moderately strong-coupling regime.

To further examine the superconducting gap symmetry in NiBi$_3$, the normalized superfluid density, $\rho_s(T)$, was calculated from the temperature-dependent magnetic penetration depth using

\begin{equation}
\rho_s(T)=\left[\frac{\lambda(0)}{\lambda(T)}\right]^2,
\end{equation}

where $\lambda(T)=\lambda(0)+\Delta\lambda(T)$ and $\lambda(0)$ is the zero-temperature penetration depth obtained from the upper critical field analysis. The temperature dependence of the normalized superfluid density is shown in the main panel of Fig 5(a).

The experimental data were analyzed using the conventional single-gap isotropic $s$-wave BCS model, where the superfluid density is calculated using Eqs.~(5) and (6). For comparison, the weak-coupling BCS curve with $2\Delta_0/k_BT_c=3.53$ is also plotted. As shown in Fig.5(a), the weak-coupling BCS model (green dotted line) deviates noticeably from the experimental data, particularly in the intermediate temperature region. In contrast, an isotropic $s$-wave fit with an adjustable gap parameter reproduces the experimental data very well over the entire temperature range. The best fit yields a superconducting gap ratio of $\frac{2\Delta_0}{k_BT_c}=4.07$, which is larger than the weak-coupling BCS value of 3.53. This result is in good agreement with the penetration depth analysis and previous study of $\mu$-SR in which they claimed the value of $\frac{2\Delta_0}{k_BT_c}=4.2$ \cite{NiBifullygap}. Our results also indicate that NiBi$_3$ is a fully gapped $s$-wave superconductor in the moderately strong-coupling regime in the dirty limit of BCS prediction.

\section{\label{sec:level6} Conclusion}
In summary, we have performed a comprehensive investigation of the superconducting properties of high-quality NiBi$_3$ single crystals using magnetization, electrical resistivity, and tunnel diode oscillator (TDO) measurements. The sharp superconducting transition ($\Delta T_c \approx 0.20$ K), together with a residual resistivity ratio (RRR) of $\sim18$, confirms the excellent crystalline quality of the samples. Magnetization measurements for both field orientations ($H \parallel b$ and $H \perp b$) reveal anisotropic upper and lower critical fields, while the extracted Ginzburg--Landau parameters establish NiBi$_3$ as an anisotropic type-II superconductor. The upper critical fields remain well below the Pauli paramagnetic limit, indicating that superconductivity is predominantly governed by orbital pair-breaking.

The superconducting gap structure has been investigated independently through the temperature dependence of both the London penetration depth and the superfluid density. The low-temperature penetration depth follows an exponential behavior and is accurately described by a single-gap isotropic $s$-wave BCS model, whereas linear and quadratic power-law models associated with nodal superconductivity fail to reproduce the experimental data. Consistently, the normalized superfluid density obtained from both the penetration depth and lower critical field measurements is well described by the single-gap $s$-wave BCS model. The penetration-depth analysis yields a superconducting gap ratio of $2\Delta_0/k_{\mathrm{B}}T_c = 4.07$, while the analysis based on $H_{c1}(T)$ gives $2\Delta_0/k_{\mathrm{B}}T_c \approx 4.12$--$4.01$. These values indicate a fully opened superconducting gap and place NiBi$_3$ in the moderately strong-coupling regime, while remaining consistent with dirty-limit BCS superconductivity.

Overall, our results provide compelling evidence that superconductivity in NiBi$_3$ is conventional, nodeless, and fully gapped with single-gap $s$-wave pairing symmetry. These findings are consistent with recent $\mu$SR studies \cite{NiBifullygap} and further establish NiBi$_3$ as an excellent platform for investigating superconductivity in Bi-based materials with strong spin--orbit coupling. The present work provides important experimental insight into the superconducting pairing mechanism of NiBi$_3$ and provides a useful experimental foundation for future investigations of its electronic structure and superconducting properties, particularly in view of the recently reported topological surface states \cite{NiBitheory,NiBiARPES}.

\section{\label{sec:level7} ACKNOWLEDGMENTS}
A. Gangwar is thankful to the Indian Institute of Technology (BHU), Varanasi – 221005, for providing a teaching assistantship. The author also gratefully acknowledges the Central Instrument Facility (CIF), IIT (BHU) for providing the necessary characterization/analytical facilities.

\bibliography{bibfile}% Produces the bibliography via BibTeX.

\end{document}